 \documentstyle[12pt]{article}
\begin{document}
 \newcommand{\be}[1]{\begin{equation}\label{#1}}
 \newcommand{\ee}{\end{equation}}
 \newcommand{\beqn}[1]{\begin{eqnarray}\label{#1}}
 \newcommand{\eeqn}{\end{eqnarray}}
\newcommand{\mat}[4]{\left(\begin{array}{cc}{#1}&{#2}\\{#3}&{#4}\end{array}
\right)}
 \newcommand{\matr}[9]{\left(\begin{array}{ccc}{#1}&{#2}&{#3}\\{#4}&{#5}&{#6}\\
{#7}&{#8}&{#9}\end{array}\right)}
 \newcommand{\eps}{\varepsilon}
 \newcommand{\Ga}{\Gamma}
 \newcommand{\la}{\lambda}
\newcommand{\ov}{\overline}
\newcommand{\mucirc}{\stackrel{\circ}{\mu}}
\newcommand{\mast}{\stackrel{\ast}{m}}
\newcommand{\meps}{\stackrel{\circ}{\eps}}
\newcommand{\mcirc}{\stackrel{\circ}{m}}
\newcommand{\mcir}{\stackrel{\circ}{M}}
\newcommand{\geqsim}{\stackrel{>}{\sim}}
\renewcommand{\thefootnote}{\fnsymbol{footnote}}

\begin{titlepage}
\begin{flushright}
INFN-FE-06-95 \\
hep-ph 9505385 \\
May  1995
\end{flushright}
\vspace{10mm}

 \begin{center}
 {\Large \bf Reconciling Present Neutrino Puzzles: \\
\vspace{3mm}
 Sterile Neutrinos as Mirror Neutrinos } \\

\vspace{1.3cm}
{\large  Zurab G. Berezhiani$^{a,b}$~ and ~
Rabindra N. Mohapatra$^c$ }
\\ [5mm]
$^a$ {\em Istituto Nazionale di Fisica Nucleare, Sezione di Ferrara,
44100 Ferrara, Italy \\ [2mm]
$^b$ Institute of Physics, Georgian Academy
of Sciences, 380077 Tbilisi, Georgia}\\ [2mm]
$^c$ {\em Department of Physics, University of Maryland, College Park,
MD 20742, U. S. A.}
\end{center}

\vspace{2mm}
\begin{abstract}


We suggest that recent neutrino puzzles that are the solar and
atmospheric neutrino deficits as well as the possible neutrino
oscillations reported by the LSND experiment and the possibility
of massive neutrinos providing the hot component of the cosmological
dark matter, can all be naturally explained
by assuming existence of a mirror world described by an
``electroweak'' gauge symmetry $[SU(2)\times U(1)]'$, with the breaking
scale larger by about factor of 30 than the scale of the standard
$SU(2)\times U(1)$ model.
An interesting aspect of this model is that the sterile neutrinos
arise from the hidden mirror sector of the theory and thus their
lightness is more natural than in the usual neutrino mass scenarios.
The needed pattern of the neutrino mass matrix in this model is obtained
by assuming a conserved ZKM-type global lepton number
$\bar L=L_e+L_\mu-L_\tau$, which is violated by Planck scale effects.
One implication of our proposal is that bulk of the dark matter in the
universe is a warm dark matter consisting of few KeV mass particles
rather than the 100 GeV range particles of the currently popular cold
dark matter scenarios.

\end{abstract}

\end{titlepage}

\renewcommand{\thefootnote}{\arabic{footnote})}
\setcounter{footnote}{0}

\newpage


The present situation in the neutrino physics is rather intriguing
but cautiously controversial. On the one hand, the direct measurements
show no evidence for any of the neutrinos to be massive,
providing only the upper bounds
$m_{\nu_e}< 4.5\,$eV ($<0.7\,$eV $[2\beta_{0\nu}]$)
$m_{\nu_\mu}< 160\,$keV and $m_{\nu_\tau}< 20\,$MeV.
The neutrinos could therefore be massless as far as these experiments
are concerned. As is well known the standard model (SM) renders its
left-handed (LH) neutrinos to be massless,
since in absence of the right-handed (RH) states the lepton
number conservation arises as an accidental global symmetry
of the theory due to the
joint requirement of gauge invariance and renormalizability.
There is of course one conceivable scenario which could lead to
neutrino masses within the minimal SM: one could imagine that
the minimal SM (or minimal $SU(5)$ GUT) is a true theory
up to the Planck scale,
however the lepton number being a global symmetry,
its conservation need not be respected by nonperturbative gravitational
effects \cite{BEG,ABS}. In that case, one can have nonrenormalizable
operators of the type $\frac{1}{M_{Pl}}l_il_jHH$ in the theory,
where $l_{1,2,3}$ are the lepton doublets and $H$ is the
Higgs doublet
(In fact, this is equivalent to the seesaw mechanism \cite{seesaw}
with the RH neutrino states having $\sim M_{Pl}$ Majorana masses.)
These can then induce small neutrino majorana masses, at most of
the order of $\mcirc =\langle H \rangle^2/M_{Pl}=3\cdot 10^{-6}\,$eV
\cite{BEG,ABS}, which value can be naturally considered
as a neutrino mass unit in the SM.

On the other hand, there are indirect ``positive''
signals for neutrino masses and mixing accumulating during past years.
In particular, if any of the following hints will prove to be true,
this would point towards neutrino masses  much larger than $\mcirc$.
This hints include:

(a) {\it The solar neutrino problem} (SNP). The solar neutrino experiments
\cite{SNP} indicate a deficiency of solar $\nu_e$ fluxes
which cannot be explained by astrophysical reasons \cite{Gianni}.
This then implies that the discrepancy between theoretical expectations
and experimental observations is due to new neutrino properties, the
most plausible one being the oscillation of $\nu_e$ into another neutrino
$\nu_x$. Interestingly, the long wavelength ``just-so'' oscillation
with $\delta m^2_{ex} \sim \mcirc ^2$ and large mixing angle, just
as the Planck scale effects predict, can provide a SNP
solution \cite{justso}. However,
the MSW oscillation \cite{MSW} appears to be the most popular and
natural solution to SNP, based on $\nu_e- \nu_x$ conversion in solar medium.
The required parameter range corresponds to
$\delta m^2_{ex}\sim 10^{-5}~\mbox{eV}^2$ and, in the preferable
``small mixing angle" scenario, $\sin^2 2\theta_{ex}\sim 10^{-3}-10^{-2}$.

(b) {\it The atmospheric neutrino problem} (ANP). There is an evidence
for a significant depletion of the atmospheric $\nu_\mu$ flux
by almost a factor of 2 \cite{ANP}. This points again to neutrino
oscillations, with a parameter range $\sin^2 2\theta_{\mu x}\sim 1$
and  $\delta m^2_{\mu x}\sim 10^{-2}~\mbox{eV}^2$.

(c) {\it Dark matter problem.} The COBE measurements of the cosmic microwave
background anisotropy suggests that cosmological dark
matter consists of two components: cold dark matter (CDM) being
a dominant component ($\Omega_{\rm CDM}\simeq 0.7$) and
hot dark matter (HDM) being a smaller admixture
($\Omega_{\rm HDM}\simeq 0.25$) \cite{HDM,HDM2}.
The latter role can be naturally played by neutrinos with mass of
few eV's.
As for the CDM candidates, several possibilities can
be envisaged. For example, in supersymmetric theories with conserved
R parity it can be provided by lightest neutralino.
However, it can be of interest to think of CDM as also consisting of
neutrinos, this time heavier (keV range), and with correspondingly
small concentration, so called warm dark matter (WDM) \cite{WDM}.

(d) {\it LSND result:}
Direct evidence of $\bar{\nu}_\mu-\bar{\nu}_e$ oscillation
from the recent Los Alamos experiment \cite{LSND}, with
$\delta m^2_{e\mu}\geq 0.3~\mbox{eV}^2$ and
$\sin^2 2\theta_{e\mu}= 10^{-3}-10^{-2}$.

Thus, these hints if proved correct
would point to the neutrino masses much larger than $\mcirc$,
which in turn would imply nontrivial physics beyond the Standard Model
at a scale below $M_{Pl}$.
In particular, this would imply that the neutrino masses are induced
from the effective operators of the type
$\frac{h_{ij}}{\Lambda} l_i l_j H H $
with the cutoff scale $\Lambda\ll M_{Pl}$.
Since this operators with $\Lambda\ll M_{Pl}$ cannot be seriously
considered as truly nonrenormalizable, they have to be effectively
generated via some renormalizable couplings of certain additional
particles - fermions or scalars, and $\Lambda$ is related to their
mass scale.  For example, in the context of seesaw
mechanism \cite{seesaw} these are the RH neutrinos.
Other neutrino mass mechanisms as are the Zee model, the heavy triplet
mechanism etc. (see e.g. \cite{book} and references therein),
also effectively reduce to the operators of these type.

A pessioptimistic\footnote{Here our points: pessimistic (Z.G.B) and
optimistic (R.N.M) did not agree.}
 approach to the situation would be to assume that
all these indications for neutrino masses
will indeed be borne out
by future experiments and one should therefore study its implications for
the texture of neutrino masses. Such a study has been carried out
in the past two years \cite{calmoh} and has led to the amazing result
that only one possible neutrino mass texture is compatible with
all the above mentioned data. It requires an extra light, sterile
neutrino $\nu_s$ beyond the three known neutrinos ($\nu_e,\nu_{\mu}$
and $\nu_{\tau}$). In this scenario the SNP is explained
as a consequence of $\nu_e$-$\nu_s$ MSW oscillation while the atmospheric
neutrino puzzle is explained by $\nu_{\mu}-\nu_{\tau}$ oscillation with
$m_{\nu_{\mu}}\simeq m_{\nu_{\tau}}\simeq 2.4$ eV. One assumes in
this scenario that $m_{\nu_{e,s}}\ll m_{\nu_{\mu,\tau}}$. The $\nu_{\mu}$
and $\nu_{\tau}$ provide the cosmological HDM and it also explains
the LSND results. The detailed mass matrix for the neutrinos in this case
is essentially in a block matrix form with each block being $2\times 2$
with a small mixing between the $\nu_s$ and $\nu_{\mu}$ to account for
the LSND result.

Indeed, the small mixing angle MSW oscillation $\nu_e -\nu_s$
is the only place where the extra sterile neutrino could show up.
Obviously, sterile neutrino cannot constitute HDM, neither it can
be applied to ANP, due to primordial nucleosynthesis bound on the
effective number of neutrino species at $t\sim 1\,$s:
$N_\nu <3.1$ \cite{NB}.
In particular, this bound implies that the oscillation of active
neutrinos into the sterile one should obey the limit
$\delta m^2 \sin^2 2\theta \leq 1.6 \cdot 10^{-6}~ \mbox{eV}^2$
\cite{Enqvist}, which excludes the $\nu_\mu-\nu_s$ oscillation as a
candidate for the ANP solution. The same bound excludes
the possibility for the SNP solution through the MSW oscillation
$\nu_e-\nu_s$ in the large mixing angle regime.

Concerning the sterile neutrinos, an immediate question that arises is
`where do they all come from, and where do they all belong': namely,
why they can be so light if their masses are allowed by the gauge
symmetry. In the SM a role of sterile neutrinos
can be played by the RH components of the usual ones.
However, in this case it is difficult to realize why they in combination
with LH neutrinos do not form the Dirac particles as heavy as
the charged fermions, or why they do not have large Majorana masses
in the spirit of the seesaw scenario. In order to render sterile
neutrinos massless, some {\em ad hoc} global symmetries should be
introduced \cite{SV,ABSZ}. The recent proposal \cite{CJS} that
the sterile neutrino can be related to axino or majorino like fields
in the context of supersymmetric theory is also based on
{\em ad hoc} constraints.

Here we suggest that the sterile neutrinos are in fact
the neutrinos of a mirror world which is the mirror duplicate
of our visible world except that its "electroweak" scale $v'$ is larger
by factor of $\zeta$ than the standard electroweak scale $v$.
It interacts with the visible world only through gravity and
possibly through some superheavy singlet scalars.
Thus, neutrinos of the mirror standard model $\nu'_{e,\mu,\tau}$
should be light by the same reason as ordinary ones $\nu'_{e,\mu,\tau}$
-- their direct mass terms are forbidden by the mirror SM gauge symmetry.
Then the SNP can be explained
through the oscillation of $\nu_e$ into its mirror partner $\nu'_e$
by means of Planck scale effects, which naturally provide
both the values of $\delta m^2$ and $\sin^2 2\theta$ in the
range needed for the small mixing angle MSW solution.
The question arises,
what is a possible role played by the two remaining mirror neutrinos.
Our answer is that in the framework presented below they have masses in
the KeV range and constitute the warm dark matter of the universe.
We also address their implications for nucleosynthesis.

While the SNP solution can be completely understood by means of
Planck scale induced MSW oscillation $\nu_e-\nu'_e$,
understanding of other puzzles asks for some non-minimal mechanisms
generating masses of the $\nu_{\mu,\tau}$ states.
%
%
We suppose that the neutrino sector of the model obeys
the separate lepton number conservation in both sectors, and assume that
these lepton numbers are broken in two stages: (A) the dominant
entries ($\sim$ eV) in the neutrino mass matrix
has origin in some intermediate scale physics which respects a
global ZKM-type lepton number ($\bar L=L_e+L_\mu-L_\tau$ in our model);
this fixes the skeleton of neutrino mass matrix with specific texture
needed for reconciling the LSND oscillation and the HDM;
(B) the conservation of this global number is violated by Planck
scale effects, which can then explain SNP and ANP.
In particular, we show that if the scale of the mirror world is some 30
times the electroweak scale, then the Planck scale induced terms
reproduce the orders of magnitudes of the parameters
required by the small mixing angle MSW scenario.

In the spirit of our hypothesis, there will be a similar pattern
for the Majorana mass matrix in the mirror sector except that the
entries should be scaled up by a factor of $\zeta^2$.
%
As far as ordinary neutrinos $\nu_{\mu,\tau}$ with $\sim$eV mass
play a HDM role, their mirror partners $\nu'_{\mu,\tau}$ being
$\zeta^2$ times heavier, for $\zeta\sim 30$ fall in keV range and thus
can constitute WDM of the universe.


\vspace{6mm}
{ \bf  Without the mirror world  }
\vspace{3mm}

Apparently, three known neutrinos
are not enough to reconcile all these hints \cite{calmoh}. In fact, the key
difficulty is related to SNP. The points (b), (c) and (d)
can be easily reconciled e.g. by assuming that in the basis of
the flavour eigenstates $\nu_{e,\mu,\tau}$ the neutrino Majorana mass
matrix has the texture obeying the $\bar L = L_e + L_\mu - L_\tau$
conservation:
\be{texture}
\hat{m}_\nu = \matr{0}{0}{a}{0}{0}{b}{a}{b}{0}
\ee
This matrix has one massless ($\nu_1$) and two massive
($\nu_{2,3}$) degenerated eigenstates,
with mass $m_{2,3}=m=(a^2+b^2)^{1/2}$. The latter can play a role of
HDM provided that $m$ is of few eV's (e.g. $m\simeq 2.4\,$eV,
according to ref. \cite{HDM2}). On the other hand,
$\nu_e$ and $\nu_\mu$ are mixed by the angle $\theta_{e\mu}$
($\tan\theta_{e\mu}=a/b$), and
$\delta m^2_{e\mu}=m^2 \simeq 6\,\mbox{eV}^2$
which can explain the LSND oscillation if
$\sin^2 2\theta_{e\mu}\simeq 2\cdot 10^{-3}$ \cite{LSND}.

Small explicit violation of $\bar L$ would induce nonzero entries in
the matrix (\ref{texture}) and thus triger the $\nu_\mu-\nu_\tau$
oscillation. For example, if the small nonzero
entry $\eps \ll m$ appears in its  (3,3) element, then we have
$\delta m^2_{\mu\tau} \approx 2\eps m$ and
almost maximal mixing, $\sin^2 2\theta_{\mu\tau}\approx 1$.
Thus, APN can be solved provided that
$\delta m^2_{\mu\tau} \simeq 10^{-2}\,\mbox{eV}^2$, which
for $m\simeq 2.4\,$eV in turn implies $\eps \simeq 2\cdot 10^{-3}\,$eV.

Thus, only the SNP remains unresolved.

\vspace{6mm}
\noindent{\bf Connecting the Two Worlds}
\vspace{4mm}

Imagine that besides the world of the usual particles
of the standard electroweak model $G=SU(2)\times U(1)$,
there exist an analogous world of particles
belonging to the mirror gauge group $G'=[SU(2)\times U(1)]'$.
In particular, the ordinary and mirror lepton states and Higgs doublets
transform as
\footnote{ All fermion states are given in the LH basis.
Quarks can be included in a strightforward way:
the ordinary quarks transform as triplets of $SU(3)_c$ while the mirror
ones are triplets of the mirror gauge group $SU(3)'_c$. Notice that
such a theory can be consistently extended to the $SU(5)\times SU(5)'$
GUT, also in the supersymmetric version. }
\beqn{leptons}
& G:~~~~~
l_{i}=\left(\begin{array}{c} \nu_i \\ e_i \end{array} \right)
\sim (\frac{1}{2},-1;0,0), ~~~~~ E^c_{i} \sim (0,2;0,0),
{}~~~~~ H\sim (\frac{1}{2},1;0,0)   \nonumber \\
& G':~~~~~
l'_{i}=\left(\begin{array}{c} \nu'_i \\ e'_i \end{array} \right)
\sim (0,0;\frac{1}{2},-1), ~~~~~E'^c_{i} \sim (0,0;0,2),
{}~~~~~ H'\sim (0,0;\frac{1}{2},1)
\eeqn
where the weak isospins $I,I'$ and hypercharges $Y,Y'$ are shown
explicitly, and index $i=1,2,3$ denotes the electron, muon and taon
families respectively.
One can impose also the invariance under discrete transformation
$P(G\leftrightarrow G')$ simoultaneously interchanging all corresponding
particles of the ordinary and mirror worlds.\footnote{ In fact,
the ordinary (LH) fermions can be interchanged with the mirror
fermions of the same helicity, as well as with their conjugated (RH) states.
In the latter case, $P$ formally plays a role of parity. }
%
We assume that $P$ is spontaneously broken, for example
by the VEV of some $P$-odd
singlet scalar $\eta$ coupled to $H$ and $H'$ \cite{P-odd}.
As a result, $H$ and $H'$ can have different nonzero VEVs:
$v'\gg v= 174\,$GeV. Hence, the mirror world is completely analogous
to ours, but with all particle masses just scaled by the factor
$\zeta=v'/v$. In particular,
for the gauge bosons we have $M_{W',Z'}=\zeta M_{W,Z}$ and,
as far as the charged leptons get masses through the
$P$-invariant Yukawa couplings
$g_i l_i E^c_i \tilde{H} + g_i l'_i E'^c_i \tilde{H}' $,
we also obtain $m'_{e,\mu,\tau}= \zeta m_{e,\mu,\tau}$.
The same applies to the ordinary and mirror quark masses.
Photons remain massless in both of these worlds \cite{mirror}.

Let us suppose that the two worlds comunicate only through gravity and
possibly also via some superheavy gauge singlet matter,
like the $P$-odd scalar $\eta$.
We do not consider the possibility of the mixed representations.
Concerning thermodynamics of the two worlds in the Early Universe,
we assume that at the inflationary reheating temperatures
they are already decoupled from each other.
If the inflaton couplings violate $P$-invariance, then one can
imagine the situation when the visible and mirror particles are
``reheated'' with different rates, so that after inflation the
effective temperatures of the ordinary and mirror thermal bathes
are different. As a result, present cosmological abundance of the
mirror particles (including lightest ``charged'' states $e',u',d'$
as well as mirror neutrinos $\nu'_{e,\mu,\tau}$ and photons $\gamma'$)
can be less than that of their visible partners.
The contribution of $\nu'$ and $\gamma'$ to
the universe expansion rate at the nucleosynthesis epoch
$t\sim 1\,$ should be very small:
$N_\nu < 3.1$ \cite{NB} which can be translated into the upper limit
on  today's abundance $r=n_{\nu'}/n_\nu < 0.04-0.02$ for $\zeta$
varied from 10 to 100.\footnote{
These estimates depend on the abundance of $\nu'$ relative to $\gamma'$
at $t\sim 1\,$s. The neutrino decoupling temperatures in the mirror and
ordinary worlds are related as $T'_D/T_D \sim \zeta^{4/3}$,
while the fermion masses are scaled by factor $\zeta$.
E.g. for $\zeta\sim 10$, $\nu'$ decouples from $\gamma'$
after the QCD$'$ phase transition (due to $P$ invariance, the `mirror'
confinement scale cannot be substantially
larger than that of the visible world:
$\Lambda'_{\rm QCD}< 1\,$GeV).
Then for their effective temperatures at $t\sim 1\,$s we obtain
$T_{\nu'}/T_{\gamma'}=(4/11)^{1/3}$.
On the contrary, for $\zeta\sim 100$ the decoupling
temperature $T'_D> 1\,$GeV and thus the contribution of light mirror
quarks $u',d'$ has also to be taken into account, which implies
$T_{\nu'}/T_{\gamma'}=(4/53)^{1/3}$. }

With the minimal particle content  (\ref{leptons}),
the neutrinos of both sectors stay massless unless one appeals to the
gravity induced Planck scale effects which
explicitly violate the global lepton number, and also can mix the
neutrino states of the two worlds \cite{ABS}.
The relevant higher order operators are ($C$ matrix is omitted):
\be{Planck}
\frac{\alpha_{ij}}{M_{Pl}} (l_{i}H)(l_{j}H) +
\frac{\alpha_{ij}}{M_{Pl}} (l'_{i}H')(l'_{j} H')
+ \frac{\beta_{ij}}{M_{Pl}} (l_iH)(l'_{j} H') + {\rm h.c.}
\ee
with the constants $\alpha,\beta\sim 1$.
More in general, an order of magnitude less values of these constants
should not come as a surprise,  while it is also possible
(e.g. in the context of string philosophy) that the actual cutoff scale
of the operators (\ref{Planck}) is order of magnitude less than $M_{Pl}$.
Hence one can let these constants to range say from $0.1$ to $10$.
Obviously, these operators generate the neutrino mass terms
in both worlds, as well as their mixing term.  Indeed,
after substituting
the VEVs of the $H,H'$ we obtain
\be{smallmass}
m_{\nu_e} = \alpha \frac{v^2}{M_{Pl}} =\alpha \mcirc ,~~~~
m_{\nu'_e} = \alpha\frac{v'^2}{M_{Pl}} = \zeta^2 \alpha\mcirc , ~~~~
m_{\nu_e\nu'_e} = \beta\frac{v v'}{M_{Pl}} =\zeta \beta \mcirc
\ee
where $\mcirc = 3\cdot 10^{-6}~ \mbox{eV}$.
As a result, $\nu_e-\nu'_e$ oscillation emerges which can
explain SNP:
\be{alter}
\delta m^2=\alpha^2 (\zeta/30)^4 \times 8\cdot 10^{-6}\,\mbox{eV}^2, ~~~~~
\sin^2 2\theta=\frac{\beta^2}{\alpha^2}(30/\zeta)^2
\times 4.5\cdot 10^{-3}
\ee
For $\alpha,\beta \sim 1$, the choice $\zeta\sim 30$
fits the  parameter range  in the regime of ``small mixing angle''
MSW solution \cite{MSW}.
More in general, taking into account the solar model uncertainties
in predicting the boron and beryllium neutrino fluxes,
the relevant parameter range can vary within
$\sin^2 2\theta= 6\cdot 10^{-4} - 2\cdot 10^{-2}$
and $\delta m^2=(4-10)\cdot 10^{-6}\,\mbox{eV}^2$ \cite{MSW}.
Then, by assuming $\alpha\sim \beta$,
the above range for $\sin^2 2\theta$ can be recovered
for $\zeta=10-100$,
while the mass difference is correlated with $\sin^2 2\theta$ as
\be{correlation}
\delta m^2\sim \alpha^2
\left(\frac{5\cdot 10^{-3}}{\sin^2 2\theta}\right)^2 \times
6\cdot 10^{-6}\,\mbox{eV}^2
\ee
which is compatible with the MSW range for a proper $\alpha$
in the interval  $0.1 -10$.

The explanation of the other neutrino puzzles needs larger
masses, which cannot be born by the Planck scale effects.
 Thus, some non-minimal neutrino mass
mechanism should be incorporated.
As we stated earlier these masses in each sector can arise only
through the effective operators respectively bilinear in $H$ and $H'$.
Such operators can be effectively induced by several mechanisms
and we will discuss one such mechanism in brief in the next section.
But on general grounds, one can write
\be{LRnu}
\frac{h_{ij}}{\Lambda} (l_iH)(l_jH) +
\frac{h_{ij}}{\Lambda'} (l'_iH')(l'_jH')
\ee
where $h_{ij}$ are some $O(1)$ `Yukawa' constants with the
$\bar L$-conserving texture (\ref{texture}), and $\Lambda\simeq \Lambda'$
are certain regulator scales.\footnote{
We will assume these scales to be independent of the VEVs of $H$ and $H'$.
For simplicity, let us take $\Lambda'\simeq \Lambda$, which seems
especially natural if $\Lambda\gg v',v$. Indeed, in order to induce
$\sim$eV entries in the mass matrix $\hat{m}_\nu$ (\ref{texture}),
$\Lambda$ should be of about $10^{13}\,$GeV. }
Then the $\nu_1\approx \nu_e$ state and its mirror partner
$\nu'_1\approx \nu'_e$ are rendered massless so that the SNP solution
can be ascribed to the Planck scale effects in a manner demonstrated
above, provided that $\zeta\sim 30$ or so.
As for the degenerated states $\nu_{2,3}$ and $\nu'_{2,3}$,
their masses are scaled as $m'\simeq \zeta^2 m$.
Therefore, for $m$ in the eV range as it is needed for explaining
the origin of HDM,
$\nu'_{2,3}$ emerge in the keV range
and thus can constitute the WDM.

Let us remark that the ANP remains unresolved unless one introduces
the explicit mechanism for $\bar L$ breaking providing nonzero
entries $\eps\sim 10^{-3}\,$eV in the mass matrix (\ref{texture}).
The Planck scale operators (\ref{Planck}) can provide at most
$\sim \mcirc$
contribution which is far below the needed value. However, in the
next section we demonstrate a model in which APN solution can be
also related to the Planck scale physics.

\vspace{6mm}
{ \bf Origin of the Dominant Neutrino Masses}
\vspace{3mm}

Let us briefly discuss how the dominant neutrino mass matrix
for each sector given in Eq. (\ref{texture}) can emerge.
As we already stated, the neutrino masses in each sector can arise only
from the operators respectively bilinear in $H$ and $H'$.
These operators can be effectively induced by several mechanisms.
The model we consider is a variant of the so called $\mu$-model
\cite{mu}. Let us assume that theory obeys conservation of individual
global lepton numbers $L_i=L_{e,\mu,\tau}$ prescribed to lepton states
in (\ref{leptons}) in obvious way, which is spontaneously violated by
the VEVs of the gauge singlet scalars with mixed lepton numbers:
$\Phi_{e\mu}$, $\Phi_{e\tau}$ and $\Phi_{\mu\tau}$,
in the sprit of refs. \cite{BH,ABSZ}. In particular, we assume that
only $\langle \Phi_{e\tau}, \Phi_{\mu\tau} \rangle\sim V$ are nonzero,
while $\langle \Phi_{e\mu} \rangle =0$. In this way, the ZKM-type
lepton number $\bar L=L_e + L_\mu - L_\tau$ remains conserved.
We also introduce the additional neutral fermions $N_i(-1),S_i(+1)$
and $N'_i(-1),S'_i(+1)$, where brackets show the individual lepton
charges $L_i$. Then the $G\times G'\times P$ invariant coupligs,
also respecting the conservation of lepton numbers, are the
following:\footnote{ Notice however that not all terms allowed by the
symmetry are introduced. Their absence can be ensured by imposing a
conservation of the individual global lepton numbers
$L_e$, $L_\mu$, $L_\tau$ and $L'_e$, $L'_\mu$, $L'_\tau$
separately in each sector. Alternatively, one
can think of the $N,S$ and $N',S'$ states as being the
$SU(2)$ and $SU(2)'$ isotriplets with zero hypercharges.
For example, in the $SU(5)\times SU(5)'$ extension of our model
these can be originated from the heavy fermionic 24-plets
in each sector. }
\be{Yuk-mu}
 {\cal L}_{Yuk}= f_i (l_i N_i H + l'_i N'_i H') +
M_i (N_i S_i + N'_i S'_i) + \lambda \Phi_{ij} (S_i S_j + S'_i S'_j)
\ee
For simplicity we assume that the Dirac mass terms are closely
degenerated: $M_{1,2,3}\sim M\gg v'$, while the Yukawa couplings
$f_i$ have  approximately the same pattern as that of
the charged fermions, say $f_3\sim 1$, $f_2\sim 3\cdot 10^{-2}$
and $f_1\sim 10^{-3}$.

Then the total mass matrices of the neutral states in each sector have
the form
\be{Mass_nu}
{\cal M} = \matr{0}{\hat{f}v}{0}{\hat{f}v}{0}{\hat{M}}
{0}{\hat{M}}{\hat{\mu}}, ~~~~~~
{\cal M}' = \matr{0}{\hat{f}v'}{0}{\hat{f}v'}{0}{\hat{M}}
{0}{\hat{M}}{\hat{\mu}}
\ee
where the Majorana mass matrix $\mu$ has a texture
conserving $\bar L= L_e + L_\mu - L_\tau$:
\be{form-mu}
\hat{\mu} = \matr{0}{0}{A}{0}{0}{B}{A}{B}{0} ,~~~~~~~
A\sim B \sim \lambda V
\ee

Let us now take into account also the effects of the possible
higher order operators explicitly violating the lepton number
conservation. Among the plethora of possible Planck scale operators
relevant are only the ones given by eq. (\ref{Planck}) and
\be{Planck-op}
\frac{\gamma_{ij}}{M_{Pl}} (S_i S_j + S'_i S'_j ) \phi^2
\ee
where the $\phi^2$ stands for gauge invariant bilinears of the scalar
fields involved and obviously it is dominated by the scalars
$\Phi_{ij}$ having the largest VEVs in our model.
Thus, the mass matrix of the $S$ states becomes
$\hat{\mu}+\hat{\gamma}\meps $, where $\meps = V^2/M_{Pl}$.
As far as the latter operator violates $\bar L$ too,
the matrix $\gamma$ in general has no zeroes.

Then, after decoupling the heavy states the
mass matrix of the light neutrinos
becomes
\be{Mtot}
\hat{m}_\nu=
\mat{\hat{s}(\hat{\mu}+\hat{\gamma}\meps )\hat{s} + \mcirc\hat{\alpha} }
{\zeta \mcirc \hat{\beta} } {\zeta \mcirc \hat{\beta}^T }
{ \zeta^2 \hat{s} (\hat{\mu}+ \hat{\gamma}\meps )\hat{s} +
\zeta^2 \mcirc\hat{\alpha} };~~~~\hat{s}=\mbox{diag}(s_1,s_2,s_3)
\ee
where
$s_{i}=f_i v/M_i$.
%
%
Clearly, in the absence of the Planck scale induced corrections
the mass matrix $\hat{m}_\nu$ of the active neutrinos has the $\bar L$
conserving texture (\ref{texture}) with $a=s_1 s_3 A$ and $b=s_2s_3 B$,
while the mass matrix of their mirror partners is just scaled by
factor of $\zeta^2$: ~$\hat{m}'_\nu=\zeta^2 \hat{m}_\nu$.
Planck scale induced corrections violate $\bar L$, giving rise to
small nonzero entries in neutrino mass matrix.
Let us consider in more details how this model solves the
present neutrino puzzles:

HDM+WDM. In order to represent the cosmological HDM component, the
mass $m$ of the degenerated states $\nu_{2,3}$ should be in the eV
range, say $m=2.4\,$eV \cite{HDM2}. Then for $\zeta\sim 30$
masses of their mirror partners $m'=\zeta^2 m$ are in the keV range
and thus can constitute cosmological WDM provided that their present
abundance relative to active neutrinos has a proper value.
This can be expressed as
%
%
\be{HWDM}
2 m = \Omega_{\rm HDM} \cdot 92 h^2 ~\mbox{eV},~~~~~
2 r m' = \Omega_{\rm WDM} \cdot 92 h^2 \mbox{eV}
\ee
where $h$ is the Hubble constant in units 100 Km s$^{-1}$ Mpc$^{-1}$.
Cosmological density of the visible baryonic matter
corresponds to $\Omega_B\simeq 0.05$. We also recall that in our model
due to different inflationary reheating,
$\Omega_{B'}< \Omega_B$
unless baryon asymmetry in the mirror world is
much larger than that of normal baryons.
Thus by taking rather conservatively $\Omega_B+\Omega_B'\leq 0.1$,
the remaining cosmological density can be shared between HDM and WDM
components as say $\Omega_{\rm HDM}\simeq 0.2$ and
$\Omega_{\rm HDM}\simeq 0.7$.
Then for the needed abundance of the mirror neutrinos relative to the
active ones we obtain
$r = \sim 3.5\cdot \zeta^{-2}$.
For $\zeta>10$ this is obviously consistent with the upper
bound  from the nucleosynthesis constraint $r<0.04-0.02$.

LSND: Oscillation $\nu_e-\nu_\mu$ occurs with
$\delta m^2_{e\mu} =m^2 \simeq 6~{\mbox eV}^2$
and  $\sin\theta_{e\mu}= f_1A/f_2B$.
Thus for our pattern of the Yukawa constants $f_i$ we obtain
$\sin^2 2\theta_{e\mu}\sim 10^{-3}$,
in the range needed for for the LSND oscillation,

ANP: The Planckian terms (\ref{Planck-op})
induce nonzero entries in the matrix $\hat{\mu}$
of the order of $\meps =V^2/M_{Pl}$. This in turn induces in
$\hat{m}_\nu$ nonzero (3,3) entry $\eps= s_3^2\gamma\meps$
which removes the $\nu_2-\nu_2$ degeneracy and
thus trigers $\nu_\mu-\nu_\tau$ oscillation with
$\sin 2\theta\approx 1$ and $\delta m^2_{\mu\tau}=2\eps m$.
Since $m= s_2s_3 \lambda V\simeq 2.4\,$eV,
then the atmospheric neutrino oscillation range requires
$\eps\simeq 2\cdot 10^{-3}\,$eV. Thus, confronting the values of
$m$ and $\eps$, we obtain the following estimates for the lepton number
breaking scale $V$ and the Dirac mass $M$:
\be{V}
V= \frac{f_2\lambda\eps}{f_3\gamma m} M_{Pl} \sim
\frac{\lambda}{\gamma}\cdot 3\cdot 10^{14}~ \mbox{GeV}, ~~~~~
M= \frac{f_2\lambda\eps}{f_3\gamma^{1/2} m}\cdot
\frac{f_3 v}{\sqrt{\eps M_{Pl}}} M_{Pl} \sim
\frac{\lambda}{\gamma^{1/2}}\cdot 10^{13}~ \mbox{GeV}
\ee
so that both scales are large for reasonable values of $\lambda$
and $\gamma$.

SNP: Obviously, the operators (\ref{Planck-op}) negligibly
contribute to the electron neutrino mass. Indeed, their contribution
to the (1,1) entry in the matrix $\hat{m}_\nu$ is
$(f_1/f_3)^2 \eps \ll \mcirc$.
Thus, for the mass terms of $\nu_e$ and $\nu'_e$
the relevant contributions come dominantly from the operators
(\ref{Planck}), leading to MSW oscillation with the parameter range
given in (\ref{alter}).

The following remark is also in order. Once the individual
global lepton numbers $L_{e,\mu,\tau}$ are broken spontaneously
down to $\bar L=L_e+L_\mu-L_\tau$, there should exist flavons
\cite{BH,ABSZ}, multiflavour analogues of the singlet
majoron \cite{Rabi}. As is known, the Planck scale effects
applied to majoron-flavons, induce their masses
$\sim (V/M)^{1/2} V$  which
overclose the universe if the massive majorons present in the today's
universe \cite{ABMS}. The estimates (\ref{V}) demonstrate, however,
that the scales $V$ and $M$ are enough large, and appearance of
the majoron after the inflationary reheating can be avoided.

In conclusion, we have shown that if the existence of the light sterile
neutrino  is to be taken seriously for solving the neutrino puzzles, then
the its lightness is most easily understood
if it is identified with the mirror neutrino.
We have then outlined a two step mechanism whereby the desired texture
needed to solve the neutrino puzzles in a unified manner can emerge.
We have given also an explicit model that can lead to the basic texture.

\vspace{4mm}
\noindent{\bf Acknowledgements.} Z.G.B thanks V. Berezinsky,
A. Dolgov, G. Fiorentini and Ursula Miscili for valuable discussions.
The work of R.N.M is supported by a grant from the National
Science Foundation under grant No. PHY9421385.

\vspace{4mm}
{\em Note added:} When our work was finished, we encountered
the paper by R. Foot and R. Volkas, preprint UM-P-95/49 (May 1995),
in which the explanation of recent neutrino puzzles is attempted
with the mirror model with exact parity. Our approach and results,
however, are drastically different from those obtained in this paper.

\end{document}